\documentclass[12pt]{article}
\usepackage{eurosym}
\usepackage{graphicx}
\usepackage[utf8]{inputenc}
\usepackage[T1]{fontenc}
\usepackage{indentfirst}
\usepackage[margin=0.6in,nomarginpar]{geometry}
\usepackage[final]{hyperref}
\usepackage{amsmath}
\usepackage{hyperref}
\usepackage{cite}
\usepackage{subcaption}
\usepackage{caption}
\usepackage{amssymb}
\usepackage{multirow}
\usepackage[table]{xcolor}
\usepackage{orcidlink}

\hypersetup{
colorlinks=true,
linkcolor=blue,
citecolor=blue,
filecolor=magenta,
urlcolor=blue
}

\begin{document}

\title{Dunkl-Schr\"{o}dinger equation with time-dependent harmonic
oscillator potential}
\author{A. Benchikha\thanks{%
benchikha4@yahoo.fr} \\
$^{1}$D\'{e}partement de EC, Facult\'{e} de SNV, Universit\'{e} Fr\`{e}res
Mentouri-Constantine 1,\\
Constantine, Algeria.\\
$^{2}$Laboratoire de Physique Math\'{e}matique et Subatomique,\\
LPMS, Universit\'{e} Fr\`{e}res Mentouri- Constantine 1, Constantine,
Algeria. \and B.Khantoul \thanks{%
boubakeur.khantoul@univ-constantine3.dz} \\
Department of Process Engineering, University of Constantine 3 - Salah
Boubnider, \\
BP: B72 Ali Mendjeli, 25000 Constantine, Algeria. \and B. Hamil\thanks{%
hamilbilel@gmail.com(Corresponding author)} \\
Laboratoire de Physique Math\'{e}matique et Subatomique,\\
LPMS, Universit\'{e} Fr\`{e}res Mentouri- Constantine 1, Constantine,
Algeria. \and B. C. L\"{u}tf\"{u}o\u{g}lu\thanks{%
bekir.lutfuoglu@uhk.cz }, \\
Department of Physics, University of Hradec Kr\'{a}lov\'{e},\\
Rokitansk\'{e}ho 62, 500 03 Hradec Kr\'{a}lov\'{e}, Czechia. }
\date{\today }
\maketitle

\begin{abstract}
In this paper, using the Lewis-Riesenfeld method, we determine the explicit
form of the wavefunctions of one- and three-dimensional harmonic oscillators
with time-dependent mass and frequency within the framework of the Dunkl
derivative, which leads to the derivation of a parity-dependent of the
invariant and auxiliary equation.
\end{abstract}



\section{Introduction}

Over the past few decades, the study of time-dependent or dynamic systems
has become increasingly important for understanding various physical
phenomena across different fields of physics \cite{Chng, Markov, Bouguerra,
Abdalla, Leach, Menouar}. The primary reasons for extensive research are the
mathematical challenges and significant applications in diverse areas of
physics, including cosmology \cite{Berger}, quantum optics \cite{Colegrave},
nanotechnology \cite{Brown}, and charged-particle beams in accelerators \cite%
{Bohn}. Significant attention has been devoted to specific problems of
time-dependent systems, including the harmonic oscillator with
time-dependent frequency and/or mass. In 1969, Lewis and Riesenfeld were the
first to present an exact solution to the harmonic oscillator problem with a
time-varying frequency\cite{Riesenfeld}. In the literature, numerous
theoretical approaches have been employed to study time-dependent quantum
mechanics, including separation of variables, perturbation theory, numerical
techniques, and invariant methods, Separation of variables decomposes the
wave function into spatial and temporal components, while perturbation
theory handles weak time dependencies by expanding the wave function in a
series. Numerical methods, like the Crank-Nicolson algorithm, approximate
solutions by discretizing time and space. Invariant methods, particularly
those using Lewis-Riesenfeld invariants. Lewis and Riesenfeld have shown
that if a system possesses an invariant $I\left( t\right) $, one can
identify a special set of eigenstates associated with this invariant
operator \cite{Riesenfeld}. The Schr\"{o}dinger's wave equation can be
written in terms of invariant operator eigenstates multiplied by an
appropriate time-dependent phase factor.

On the other hand, in various fields of physics and mathematics, quantum
algebras, groups, and deformations have been the subject of extensive
research \cite{Chung, Chaichian, Crnugelj}. In recent years there has been
particular interest in Wigner-Dunkl quantum mechanics \cite{Dong, WSC2023,
Junker, Quesne}, which dates back to the pioneering work of Wigner and Yang 
\cite{Wigner, Yang}, in which they explored the determination of commutation
relations from equations of motion. In \cite{Dunkl}, mathematician Charles
Dunkl examined differential-difference and reflection groups along with
their symmetries and expanded upon Yang's suggestion by introducing the
differential-difference operator. Subsequently, the Dunkl derivative has
been widely applied in various fields of mathematics \cite{Dunkl1, Xu} and
physics \cite{Kakei, Hikami, Lapointe, Mik1, Mik2, Plyushchay, Horvathy,
Luo, Schulze, Schulze1, Schulze2}. In this context, he introduced the Dunkl
derivative with an additional term that distinguishes it from the standard
derivative in the following form 
\begin{equation}
\hat{D}_{j}=\frac{\partial }{\partial x_{j}}+\frac{\mu _{j}}{x_{j}}\left( 1-%
\hat{R}_{j}\right) ,  \label{P1}
\end{equation}%
with the Wigner parameter, $\mu $, and the reflection operator, $\hat{R}$,
which satisfies%
\begin{equation}
\hat{R}_{j}f\left( x_{j}\right) =f\left( -x_{j}\right), \quad \quad \hat{R}%
_{i} \hat{R}_{j}=\hat{R}_{j}\hat{R}_{i}, \quad \quad \hat{R}%
_{i}x_{j}=-\delta_{ij}x_{j}\hat{R}_{i},\quad \quad \hat{R}_{i}\frac{\partial 
}{\partial x_{j}}=-\delta _{ij}\frac{\partial }{\partial x_{j}}\hat{R}_{i}.
\label{P2}
\end{equation}
The effect of the reflection operator within the Dunkl operator produces
different effects depending on whether the function it operates on is even
or odd. Because of this perspective, the Dunkl operator has been
increasingly used, especially in the last decade, as a replacement for the
conventional partial derivative operator in quantum mechanical systems \cite%
{G1, G2, G3, G4, Ramirez1, Sargol, Ojeda, Mota1, Bilel3, BilalEPL, Mota2,
Mota3, Bilel1, Merad, Genest2015Coulomb, Ramirez2, Ghaz, Ghaz1, Kim, Ghaz2,
MotGul, MotMie, Jang, ChungDunkl, Chungrev, Chungqm, Marcelo, superstat,
Bilel2, BEC1, BEC2, BEC3, BEC4, DRelat1, BCLind, Sam1}.

To the best of our knowledge, time-dependent systems have not been
extensively studied in the framework of the Dunkl formalism. In order to
fill this gap, in this manuscript we consider the time-dependent harmonic
oscillator both in one dimension and in three dimensions. To do this, we
have structured this manuscript as follows: In Section \ref{sec2}, we derive
solutions for the time-dependent Dunkl-Schr\"{o}dinger equation in one
dimension, considering a harmonic oscillator with time-varying mass and
frequency. We employ invariant methods, expressing solutions in terms of
generalized Laguerre polynomials, and subsequently compute the quantum
phase. In Section \ref{sec3}, we extend our analysis to the
three-dimensional case in spherical coordinates, still considering a
time-dependent harmonic oscillator with time-varying mass and frequency. We
again utilize invariant methods and compute the resulting real phase.
Finally, we give a short conclusion.

\newpage

\section{Time-dependent harmonic oscillator in one dimension}

\label{sec2}

In this manuscript, we solve the time-dependent Schr\"{o}dinger equation of
the form 
\begin{equation}
H\left( t\right) \psi \left( x,t\right) =i\hbar \frac{\partial }{\partial t}%
\psi \left( x,t\right) ,  \label{1}
\end{equation}%
to describe the behavior of a quantum system whose Hamiltonian explicitly
varies with time. Here, we use the Lewis-Riesenfeld method, which provides a
technique for solving the time-dependent Schr\"{o}dinger equation when a
certain condition is met. Basically, this condition involves the existence
of an invariant Hermitian operator, $I\left( t\right) $, which satisfies 
\begin{equation}
\frac{dI\left( t\right) }{dt}=\frac{\partial I\left( t\right) }{\partial t}+%
\frac{1}{i\hbar }\left[ I\left( t\right) ,H\left( t\right) \right] =0,
\label{P9}
\end{equation}%
such that a particular solution of Eq. (\ref{1}), $\psi \left( x,t\right) $,
appears in the form of 
\begin{equation}
\psi \left( x,t\right) =e^{i\eta \left( t\right) }\Phi \left( x,t\right) ,
\label{ph}
\end{equation}%
where $\Phi _{n}\left( x,t\right) $ are the eigenfunctions of the Hermitian
operator determined from 
\begin{equation}
I\left( t\right) \Phi \left( x,t\right) =\lambda \Phi \left( x,t\right) .
\end{equation}%
Here, $\lambda _{n}$ represents the time-independent eigenvalues of $I\left(
t\right) $, and $\eta \left( t\right) $ are phase functions, derived from
the equation%
\begin{equation}
\hbar \frac{d}{dt}\eta \left( t\right) =\left\langle \Phi \left( x,t\right)
\right\vert \left( i\hbar \frac{\partial }{\partial t}-H\left( t\right)
\right) \left\vert \Phi \left( x,t\right) \right\rangle .  \label{12}
\end{equation}%
We then replace the ordinary derivatives with the Dunkl ones, and for a
time-dependent one-dimensional harmonic oscillator we get the Dunkl
Hamiltonian: 
\begin{equation}
H=-\frac{\hbar ^{2}}{2M(t)}\left( \frac{\partial ^{2}}{\partial x^{2}}+\frac{%
2\mu }{x}-\frac{\mu \left( 1-R\right) }{x^{2}}\right) +\frac{1}{2}M(t)\omega
^{2}\left( t\right) x^{2}.  \label{5}
\end{equation}%
We note that the Dunkl-Hamiltonian and the reflection operator commute with
each other, so they should have common eigenfunctions and thus can be
diagonalised simultaneously. This fact allows us to choose the eigenfunction 
$\psi \left( x,t\right) $ with a definite parity: $R\psi \left( x,t\right)
=s\psi \left( x,t\right) $, where $s=\pm $. Consequently, the Dunkl-Schr\"{o}%
dinger equation takes the following form: 
\begin{equation}
\left[ \frac{\mathcal{P}^{2}}{2M(t)}+\frac{\hbar ^{2}\left( \nu
^{2}-1/4\right) }{2M\left( t\right) x^{2}}+\frac{1}{2}M(t)\omega ^{2}\left(
t\right) x^{2}\right] \psi ^{s}\left( x,t\right) =i\hbar \frac{\partial \psi
^{s}\left( x,t\right) }{\partial t},  \label{q1}
\end{equation}%
where $\mathcal{P}=\frac{\hbar }{i}\left( \frac{\partial }{\partial x}+\frac{%
\mu }{x}\right) $, and $\nu =\mu -\frac{s}{2}.$

It is worth emphasizing that this transformation simplifies the equation,
allowing us to concentrate on solving for $\psi^{s}\left( x,t\right) $ and
ultimately to understand the behavior of the system under the time-dependent
harmonic potential.

\subsection{Derivation of the Dunkl-invariant operator}

In the latter formalism, the exact invariant operator corresponds to the
quantum system described by Eq. (\ref{q1}), so it has to be called the Dunkl
exact invariant operator since it involves the Wigner constant and the
parity factor. To derive $I^{s}\left( t\right) $, we assume that it has the
following form:%
\begin{equation}
I^{s}=\frac{1}{2}\Big(\alpha \left( t\right) T_{1}+\beta \left( t\right)
T_{2}+\gamma \left( t\right) T_{3}\Big),  \label{q2}
\end{equation}%
with real functions $\alpha ,$ $\beta ,$ and $\gamma $ and generators $%
\left\{ T_{1},T_{2},T_{3}\right\} $ of the forms: 
\begin{equation}
T_{1}=\mathcal{P}^{2}+\frac{\hbar ^{2}\left( \nu ^{2}-1/4\right) }{x^{2}}%
,\quad \quad T_{2}=x^{2},\quad \quad T_{3}=x\mathcal{P}+\mathcal{P}x,
\label{q3}
\end{equation}%
which obey the following commutations 
\begin{equation}
\left[ T_{1},T_{2}\right] =-2i\hbar T_{3},\text{ \ \ }\left[ T_{2},T_{3}%
\right] =4i\hbar T_{2},\text{ \ \ }\left[ T_{1},T_{3}\right] =-4i\hbar T_{1}.
\label{q4}
\end{equation}%
After solving the Eq. (\ref{P9}), we find that the real functions must
satisfy the following system of coupled first-order linear differential
equations: 
\begin{eqnarray}
\alpha  &=&\rho ^{2},  \label{i1} \\
\beta  &=&\frac{1}{\rho ^{2}}+M^{2}\dot{\rho}^{2},  \label{i2} \\
\gamma  &=&-M\rho \dot{\rho}.  \label{i3}
\end{eqnarray}%
Using Eqs. (\ref{i1}), (\ref{i2}) and (\ref{i3}) in Eq. (\ref{q2}), we
obtain the Dunkl exact invariant as 
\begin{equation}
I^{s}=\frac{1}{2}\left[ \rho ^{2}\left( \mathcal{P}^{2}+\frac{\hbar
^{2}\left( \nu ^{2}-1/4\right) }{x^{2}}\right) +\left( \frac{1}{\rho ^{2}}%
+M^{2}\dot{\rho}^{2}\right) x^{2}-\rho \dot{\rho}M\left( x\mathcal{P}+%
\mathcal{P}x\right) \right] ,  \label{q6}
\end{equation}%
where $\rho $ is a real parameter satisfying the Ermakov-Pinney equation 
\cite{Ermakov,Pinney}:%
\begin{equation}
\ddot{\rho}+\frac{\dot{M}}{M}\dot{\rho}+\omega ^{2}\rho =\frac{1}{M^{2}\rho
^{3}}.  \label{q7}
\end{equation}%
To derive the explicit form of the wavefunctions, we have to identify a
preferred basis of eigenstates for the Dunkl invariant by solving the
following eigenvalue equation. 
\begin{equation}
I^{s}\Phi _{n}^{s}=\lambda _{n}^{^{s}}\Phi _{n}^{s}.
\end{equation}%
Now, we introduce a unitary transformation 
\begin{equation}
\varphi _{n}^{s}\left( x\right) =U\Phi _{n}^{s}\left( x\right) =\exp \left( 
\frac{iM\dot{\rho}}{2\hbar \rho }x^{2}\right) \Phi _{n}^{s}\left( x\right) ,
\label{uni}
\end{equation}%
which transforms the operator $I^{s}$ to $\mathcal{I}^{s}$: 
\begin{equation}
\mathcal{I}^{s}=UI^{s}U^{\dag }=\frac{1}{2}\left[ \rho ^{2}\left( \mathcal{P}%
^{2}+\frac{\hbar ^{2}\left( \nu ^{2}-1/4\right) }{x^{2}}\right) +\frac{1}{%
\rho ^{2}}x^{2}\right] .  \label{q8}
\end{equation}%
Then, using the substitution $\mathcal{I}^{s}\varphi ^{s}\left( x\right)
=\lambda _{n}^{^{s}}\varphi ^{s}\left( x\right) $ with $\varphi ^{s}\left(
x\right) =\left\vert x\right\vert ^{-\mu }\phi ^{s}\left( x\right) $ and the
new variable $y=\frac{x}{\rho }$, we get the eigenvalue equation for the
operator $\mathcal{I}^{s}$ 
\begin{equation}
\frac{\partial ^{2}}{\partial y^{2}}\phi ^{s}\left( x\right) +\frac{1}{\hbar
^{2}}\left[ 2\lambda -\frac{\hbar ^{2}\left( \nu ^{2}-1/4\right) }{y^{2}}%
-y^{2}\right] \phi ^{s}\left( x\right) =0.  \label{q9}
\end{equation}%
By putting the parameter $\lambda _{n}^{s}=\hbar \left( 2n+1+\nu \right) $,
with $n=0,1,2...,$ we express the solution of the above equation in terms of
generalized Laguerre polynomials \cite{grad}, 
\begin{equation}
\phi _{n}^{s}\left( x\right) =\mathbf{N}_{s,n}\text{ }y^{\mu +\frac{1-s}{2}%
}e^{-\frac{y^{2}}{2\hbar }}L_{n}^{\mu -s/2}\left( \frac{y^{2}}{\hbar }%
\right) ,  \label{q10}
\end{equation}%
with the normalization constant, $\mathbf{N}_{s,n}$. Finally, combining $%
\varphi ^{s}\left( x\right) $ with the wave function $\phi ^{s}\left(
x,t\right) $ and using the variable transformation $y=\frac{x}{\rho }$, we
get the wave function as 
\begin{equation}
\Phi _{n}^{s}\left( x,t\right) =\mathbf{N}_{s,n}\frac{\left\vert
x\right\vert ^{-\mu }x^{\mu +\frac{1-s}{2}}}{\rho ^{\mu +\frac{1-s}{2}}}%
L_{n}^{\mu -s/2}\left( \frac{x^{2}}{\hbar \rho ^{2}}\right) \exp \left[
\left( iM\rho \dot{\rho}-1\right) \frac{x^{2}}{2\hbar \rho ^{2}}\right] .
\label{q12}
\end{equation}

\subsection{Quantum phase and total wave function}

We now focus on deriving the quantum phase using Eq. (\ref{12}). To
determine the phase function, $\eta _{n}^{s}\left( t\right) $, we use Eqs. (%
\ref{q1}) and (\ref{uni}) in Eq. (\ref{12}). This gives us 
\begin{equation}
\dot{\eta}_{n}^{s}\left( t\right) =-\frac{\lambda _{n}^{s}}{\hbar M\rho ^{2}}%
.
\end{equation}
Then, we integrate it over time 
\begin{equation}
\eta _{n}^{s}\left( t\right) =-\left( 2n+1+\mu -\frac{s}{2}\right)
\int_{0}^{t}\frac{dt^{\prime }}{M\left( t^{\prime }\right) \rho
\left(t^{\prime }\right) ^{2}}.  \label{pha}
\end{equation}%
We observe that the quantum phase varies due to the parity. Utilizing Eqs. (%
\ref{ph}), (\ref{q12}) and (\ref{pha}) we find the normalized wavefunctions
with the corresponding eigenvalues for both cases as follows:

\begin{itemize}
\item Even parity case: \bigskip $s=+1,$

the wavefunction 
\begin{eqnarray}
\psi _{n}^{+}\left( x,t\right) &=&\sqrt{\tfrac{n!}{\hbar ^{\mu +\frac{1}{2}%
}\Gamma \left( \mu +n+\frac{1}{2}\right) }}\rho ^{-\left( \mu +\frac{1}{2}%
\right) }\left\vert x\right\vert ^{-\mu }x^{\mu }L_{n}^{\mu -\frac{1}{2}%
}\left( \frac{x^{2}}{\hbar \rho ^{2}}\right)  \notag \\
&\times&\exp \Bigg[ \left( iM\left( t\right) \rho \dot{\rho}-1\right) \frac{%
x^{2}}{2\hbar \rho ^{2}}-i\left( 2n+\mu +\frac{1}{2}\right) \int_{0}^{t}%
\frac{dt^{\prime }}{M\left( t^{\prime }\right) \rho \left( t^{\prime
}\right) ^{2}}\Bigg] ,
\end{eqnarray}
and the eigenvalue 
\begin{equation}
\lambda _{n}^{+}=2n+\mu +\frac{1}{2}.
\end{equation}

\item Odd parity: $s=-1,$

the wavefunction 
\begin{eqnarray}
\psi _{n}^{-}\left( x,t\right) &=&\sqrt{\tfrac{n!}{\hbar ^{\mu +\frac{3}{2}%
}\Gamma \left( \mu +n+\frac{3}{2}\right) }}\rho ^{-\left( \mu +\frac{3}{2}%
\right) }x^{\mu +1}\left\vert x\right\vert ^{-\mu }L_{n}^{\mu +\frac{1}{2}%
}\left( \frac{x^{2}}{\hbar \rho ^{2}}\right) \\
&\times& \exp \Bigg[ \left( iM\left( t\right) \rho \dot{\rho}-1\right) \frac{%
x^{2}}{2\hbar \rho ^{2}}-i\left( 2n+\mu +\frac{3}{2}\right) \int_{0}^{t}%
\frac{dt^{\prime }}{M\left( t^{\prime }\right) \rho \left( t^{\prime
}\right) ^{2}}\Bigg] ,
\end{eqnarray}

and the eigenvalue, 
\begin{equation}
\lambda _{n}^{-}=2n+\mu +\frac{3}{2}.
\end{equation}
\end{itemize}

We note that in both cases, the eigenvalues depend on the Wigner constant.

\newpage

\section{ Time-dependent isotropic harmonic oscillator in three dimensions}

\label{sec3}

In this section, we study the non-relativistic time-dependent isotropic
harmonic oscillator problem within the Dunkl formalism in three dimensions.
In this case, we have to consider the time-dependent Dunkl-Schr\"{o}dinger
equation 
\begin{equation}
\mathcal{H}\Psi =i\hbar \frac{\partial }{\partial t}\Psi ,  \label{13}
\end{equation}%
with the following Dunkl-Hamiltonian operator 
\begin{equation}
\mathcal{H}=-\frac{\hbar ^{2}}{2M(t)}\bigg(\hat{D}_{1}^{2}+\hat{D}_{2}^{2} + 
\hat{D}_{3}^{3}\bigg) +\frac{M(t)}{2}\omega ^{2}\left( t\right) \Big(%
x_{1}^{2}+x_{2}^{2}+x_{3}^{2}\Big) ,
\end{equation}%
where 
\begin{eqnarray}
\hat{D}_{i}^{2}=\frac{d^{2}}{dx_i^{2}}+\frac{2\mu_i }{x_i}\frac{d}{dx_i}-%
\frac{\mu_i }{x_i^{2}}\left( 1-R_i\right).
\end{eqnarray}
To find the solution of Eq. (\ref{13}), it is preferable to employ the
spherical coordinate 
\begin{equation}
x_{1}=r\cos \varphi \sin \theta ,\text{ \ \ }x_{2}=r\sin \varphi \sin \theta
,\text{ \ \ }x_{3}=r\cos \varphi,
\end{equation}%
and so the Hamiltonian becomes 
\begin{equation}
\mathcal{H}=\frac{1}{2M(t)}\mathbf{p}^{2}+\frac{\delta \left( \delta
-1\right) }{M(t)r^{2}}+\frac{1}{2}M(t)\omega ^{2}\left( t\right) r^{2},
\label{hamilto}
\end{equation}
with the momentum operator, 
\begin{equation}
\mathbf{p}^{2}=\mathcal{P}_{r,\delta }^{2}+\frac{L_{D}^{2}}{r^{2}}.
\label{Dunkl momentum}
\end{equation}
Here, the operator $\mathcal{P}_{r,\delta }$ (\ref{Dunkl momentum}) is
defined as 
\begin{equation}
\mathcal{P}_{r,\delta }=\frac{\hbar }{i}\left( \frac{\partial }{\partial r}+%
\frac{\delta }{r}\right),
\end{equation}
while the Dunkl-angular momentum operator is given by 
\begin{eqnarray}
L_{D}^{2} &=&-\hbar ^{2}\Bigg[ \frac{\partial ^{2}}{\partial \theta ^{2}}%
+2\left( \left( \frac{1}{2}+\mu _{1}+\mu _{2}\right) \cot \theta -\mu
_{3}\tan \theta \right) \frac{\partial }{\partial \theta }-\frac{\mu _{3}}{%
\cos ^{2}\theta }\left( 1-R_{3}\right) \   \notag \\
&+&\frac{1}{\sin ^{2}\theta }\left( \frac{\partial ^{2}}{\partial \varphi
^{2}}+2\left( \mu _{2}\cot \varphi -\mu _{1}\tan \varphi \right) \frac{%
\partial }{\partial \theta }-\frac{\mu _{1}\left( 1-R_{1}\right) }{\cos
^{2}\varphi }-\frac{\mu _{2}\left( 1-R_{2}\right) }{\sin ^{2}\varphi }%
\right) \Bigg] ,  \label{angulair operator}
\end{eqnarray}
with 
\begin{eqnarray}
\delta =\mu _{1}+\mu _{2}+\mu _{3}+1, \quad \text{and} \quad \left[ r,%
\mathcal{P} _{r,\delta \text{ }}\right] =i\hbar.
\end{eqnarray}
We then establish the Dunkl exact invariant for Eq. (\ref{hamilto}) in the
same form as proposed in the previous section with Eq. (\ref{q2}), but with
new generators 
\begin{equation}
\mathcal{T}_{1}=\mathbf{p}^{2}+\frac{\hbar ^{2}\delta \left( \delta-1\right) 
}{r^{2}},\quad \quad \mathcal{T}_{2}=r^{2},\quad \quad \mathcal{T}_{3}=r%
\mathcal{P}_{r,\delta \text{ }}+\mathcal{P}_{r,\delta \text{}}r,
\label{Generatore 3d}
\end{equation}
which obey the commutation relations given in Eq. (\ref{q4}). We then use
Eqs. (\ref{Generatore 3d}) and (\ref{q4}) and obtain the explicit expression
of the invariant $I_{3d}^{s}$ 
\begin{equation}
I_{3d}^{s}=\frac{1}{2}\left[ \left( \frac{1}{\rho ^{2}}+M^{2}\dot{\rho}%
^{2}\right) r^{2}+\rho ^{2}\left( \mathbf{p}^{2}+\frac{\hbar ^{2}\delta
\left( \delta -1\right) }{r^{2}}\right) -\rho \dot{\rho}M\left( r\mathcal{P}%
_{r,\delta \text{ }}+\mathcal{P}_{r,\delta \text{ }}r\right) \right] .
\label{invariant 3d}
\end{equation}
where $\rho $ should satisfy Ermakov-Pinney equation.

Now, we apply the eigenvalue equation $I_{3d}\Phi =\varepsilon $ $\Phi $,
and use a unitary transformation. 
\begin{equation}
\Phi =S\text{ }\digamma =e^{\frac{iM\dot{\rho}}{2\hbar \rho }r^{2}}\digamma .
\end{equation}%
Consequently, the invariant $I_{3d}^{s}$ turns into $I_{3d}^{s}=S^{+}%
\digamma S=\mathcal{I}_{3d}^{s}$ which has form 
\begin{equation}
\mathcal{I}_{3d}^{s}=\frac{1}{2}\left[ \rho ^{2}\left( \mathbf{p}^{2}+\frac{%
\hbar ^{2}\delta \left( \delta -1\right) }{r^{2}}\right) +\frac{1}{\rho ^{2}}%
r^{2}\right] ,  \label{transf 1}
\end{equation}%
and therefore the eigenvalue equation changes to 
\begin{equation}
\mathcal{I}_{3d}^{s}\digamma =\varepsilon _{n_{r}}\text{ }\digamma .
\label{253}
\end{equation}%
Before solving Eq.(\ref{253}), we have to clarify how the reflection
operators affect the spherical wave functions. 
\begin{equation}
R_{1}f\left( r,\theta ,\varphi \right) =f\left( r,\theta ,\pi -\varphi
\right) ,\quad \quad R_{2}f\left( r,\theta ,\varphi \right) =f\left(
r,\theta ,-\varphi \right) ,\quad \quad R_{3}f\left( r,\theta ,\varphi
\right) =f\left( r,\pi -\theta ,\varphi \right) .
\end{equation}%
Next, we assume $\digamma =r^{-\delta }\Upsilon \left( r\right) \Theta
\left( \theta \right) \Phi \left( \varphi \right) $ and substitute it in Eq.
(\ref{253}). This allows us to separate Eq. (\ref{253}) into three ordinary
differential equations, 
\begin{eqnarray}
\left\{ \frac{\partial ^{2}}{\partial \varkappa ^{2}}+\frac{1}{\hbar ^{2}}%
\left[ 2\varepsilon _{n_{r}}-\varkappa ^{2}-\hbar ^{2}\frac{\delta \left(
\delta -1\right) +q^{2}}{\varkappa ^{2}}\right] \right\} \Upsilon \left(
\varkappa \right)  &=&0,  \label{radial2} \\
\left\{ \frac{\partial ^{2}}{\partial \theta ^{2}}+2\left[ \left( \frac{1}{2}%
+\mu _{1}+\mu _{2}\right) \cot \theta -\mu _{3}\tan \theta \right] \frac{%
\partial }{\partial \theta }-\frac{\mu _{3}}{\cos ^{2}\theta }\left(
1-R_{3}\right) +q^{2}\right\} \Theta \left( \theta \right)  &=&0,
\label{angular 1} \\
\left\{ \frac{\partial ^{2}}{\partial \varphi ^{2}}+2\left( \mu _{2}\cot
\varphi -\mu _{1}\tan \varphi \right) \frac{\partial }{\partial \theta }-%
\frac{\mu _{1}}{\cos ^{2}\varphi }\left( 1-R_{1}\right) -\frac{\mu _{1}}{%
\cos ^{2}\varphi }\left( 1-R_{2}\right) +k^{2}\right\} \Phi \left( \varphi
\right)  &=&0.  \label{angulair 2}
\end{eqnarray}%
Here $\varkappa =\frac{r}{\rho }$, with $k$ and $q$\ being two separation
constants. We note that Eqs. (\ref{angular 1}) and (\ref{angulair 2})
resemble those found in the study of the three-dimensional Dunkl-harmonic
oscillator problem \cite{G4}. Consequently, we express their solutions using
Jacobi polynomials characterized by the parity quantum numbers $s_{1}$, $%
s_{2}$, $s_{3}$, 
\begin{eqnarray}
\Phi _{m}^{s_{1},s_{2}}\left( \varphi \right)  &=&\mathcal{C}_{\varphi }\cos
^{s_{1}}\left( \varphi \right) \sin ^{s_{2}}\left( \varphi \right) 
\boldsymbol{P}_{m-\frac{s_{1}+s_{2}}{2}}^{\mu _{2}+s_{2}-1/2,\mu
_{1}+s_{1}-1/2}\left( \cos 2\varphi \right) ,  \label{Solu1} \\
\Theta \left( \theta \right)  &=&\mathcal{C}_{\theta }\cos ^{s_{3}}\left(
\theta \right) \sin ^{2m}\left( \theta \right) \boldsymbol{P}_{l-\frac{s_{3}%
}{2}}^{\left( 2m+\mu _{1}+\mu _{2},\mu _{3}+s_{3}-\frac{1}{2}\right) }\left(
\cos \left( 2\theta \right) \right) ,  \label{solu2}
\end{eqnarray}%
where $\mathcal{C}_{\varphi }$ and $\mathcal{C}_{\theta }$ are the
normalization constants, Here, the separation constants have to satisfy the
following forms: 
\begin{eqnarray}
k^{2} &=&4m\left( m+\mu _{1}+\mu _{2}\right) ,  \label{sepa1} \\
q^{2} &=&4\left( l+m\right) \left( l+m+\mu _{1}+\mu _{2}+\mu _{3}+\frac{1}{2}%
\right) .  \label{sepa2}
\end{eqnarray}%
It is worth noting that $m$ is a positive half-integer if $s_{1}s_{2}=-1$,
otherwise, it is a non-negative integer. In addition, the quantum number $l$
takes non-negative integer values if $s_{3}=+1$, and positive half-integer
values if $s_{3}=-1$. Next, we proceed with the radial component given in
Eq. (\ref{radial2}). The radial equation resembles the scenario discussed in %
\ref{sec2}, leading to identical outcomes as described in Eq. (\ref{q9}).
The solution of Eq. (\ref{radial2}) is 
\begin{equation}
\Upsilon _{n_{r},m,l}\left( \varkappa \right) =\mathcal{C}_{r}\varkappa
^{\sigma +\frac{1}{2}}e^{-\frac{1}{2\hbar }\varkappa ^{2}}L_{n_{r}}^{\sigma
}\left( \frac{\varkappa ^{2}}{\hbar }\right) ,
\end{equation}%
with the constant eigenvalues%
\begin{equation}
\varepsilon _{n_{r},l,m}=\hbar \left( 2n_{r}+\sigma +1\right) ,\quad \quad
n_{r}=0,1,\cdots .  \label{eigenvalue}
\end{equation}%
Here, $\mathcal{C}_{r}$ denotes the normalization constant, and 
\begin{equation}
\sigma =\sqrt{\frac{1}{4}+\delta \left( \delta -1\right) +4\left( l+m\right)
\left( l+m+\mu _{1}+\mu _{2}+\mu _{3}+\frac{1}{2}\right) }.
\end{equation}%
In the end, we are left with the challenge of determining the phases $\eta
\left( t\right) $ that satisfy Eq. (\ref{ph}). By applying the unitary
transformation $S$, Eq. (\ref{ph}) transforms into 
\begin{equation}
\dot{\eta}_{n_{r},l,m}\left( t\right) =-\frac{\varepsilon _{n_{r},l,m}}{%
M\rho ^{2}}.  \label{ph1}
\end{equation}%
Therefore, the quantum phase reads:%
\begin{equation}
\eta _{n_{r},l,m}\left( t\right) =-\left( 2n_{r}+\sigma +1\right)
\int_{0}^{t}\frac{dt^{\prime }}{M\left( t^{\prime }\right) \rho \left(
t^{\prime }\right) ^{2}}.
\end{equation}%
Before ending this section, it should be mentioned that the normalization
constants $\mathcal{C}_{r}$, $\mathcal{C}_{\theta }$ and $\mathcal{C}%
_{\varphi }$ can be calculated from the following normalization condition 
{\small 
\begin{eqnarray}
&&\int_{0}^{+\infty }\int_{0}^{\pi }\int_{0}^{2\pi }r^{2+2\mu _{1}+2\mu
_{2}+2\mu _{3}}\left\vert \sin \theta \right\vert ^{2\mu _{2}+2\mu
_{1}}\left\vert \cos \theta \right\vert ^{2\mu _{3}}\left\vert \sin \varphi
\right\vert ^{2\mu _{2}}\left\vert \cos \varphi \right\vert ^{2\mu _{1}}\sin
\theta \psi _{n_{r}^{\prime },l^{\prime },m^{\prime }}^{\ast \left(
s_{1}^{\prime },s_{2}^{\prime },s_{3}^{\prime }\right) }\psi
_{n_{r},l,m}^{\left( s_{1},s_{2},s_{3}\right) }drd\theta d\varphi   \notag \\
&=&\delta _{n_{r,n_{r}}^{\prime }}\delta _{l,l^{\prime }}\delta
_{m,m^{\prime }}\delta _{s_{1},s_{1}^{\prime }}\delta _{s_{2},s_{2}^{\prime
}}\delta _{s_{3},s_{3}^{\prime }}.
\end{eqnarray}%
}

\section{\protect\normalsize Conclusions}

{\normalsize Especially in the last decade, studies proposing to replace the
ordinary derivative with the Dunkl derivative have provided important
insights into our understanding of physical systems. This is essentially
based on the fact that the Dunkl derivative adds simultaneous
mirror-symmetric solutions to the solution of the systems under
consideration. In addition, the free parameter it contains can be used to
obtain a better fit between experimental and theoretically proposed results.
Bearing all these facts in mind we have revisited the time-dependent quantum
harmonic oscillator problem by considering a time-varying mass and frequency
within the Dunkl formalism. First, we treated the problem in one dimension
and constructed the corresponding Dunkl-Schr\"{o}dinger equation. We then
used the Lewis-Riesfeld method and determined the exact invariant operator.
We then derived the parity-dependent total wave function solutions, the
eigenvalues and the quantum phase. Next, we considered the isotropic
harmonic oscillator in three dimensions and solved the Dunkl-Schr\"{o}dinger
equation using the Lewis-Riesfeld method in spherical coordinates for
convenience. convenience. We found that not only the radial equation but
also the solutions of the angular equations can be classified as mirror
symmetric solutions. }

{\normalsize 
}

{\normalsize 
}

\section*{\protect\normalsize Data Availability Statements}

{\normalsize The authors declare that the data supporting the findings of
this study are available within the article. }

\end{document}